\documentclass{cernrep}
\usepackage{texnames}
\usepackage[T1]{fontenc}
\usepackage{url}
\usepackage{amsmath}
\usepackage{graphicx}

\begin{document}
\title{Practical Statistics}
 
\author{L.~Lyons}

\institute{Blackett Lab., Imperial College, London, UK and Particle Physics, Oxford, UK}

\maketitle 





  
  


\begin{abstract}
Accelerators and detectors are expensive, both in terms of money and human effort. It is thus important
to invest effort in performing a good statistical analysis of the data, in order to extract the best information
from it. This series of five lectures deals with practical aspects of statistical issues that arise in typical 
High Energy Physics analyses.
\end{abstract}

\section{Outline}


 This series of five lectures deals with practical aspects of statistical issues that arise in typical 
High Energy Physics analyses. The topics are:
\begin{itemize}
\item{Introduction. This is largely a reminder of topics which you should have encountered as 
undergraduates. Some of them are looked at in novel ways, and will hopefully provide new insights.}
\item{Least Squares and Likelihoods. We deal with two different methods for parameter determination.
Least Squares is also useful for Goodness of Fit testing, while likelihood ratios play a crucial role in
choosing between two hypotheses.}
\item{Bayes and Frequentism. These are two fundamental and very different approaches to statistical
searches. They disagree even in their views on `What is probability?'}
\item{Searches for New Physics. Many statistical issues arise in searches for New Physics. These may 
result in discovery claims, or alternatively in exclusion of theoretical models in some region of their
parameter space (e.g. mass ranges).} 
\item{Learning to love the covariance matrix. This is relevant for dealing with the possible correlations
between uncertainties on two or more quantities. The covariance matrix takes care of all these
correlations, so that you do not have to worry about each situation separately.
This was an unscheduled lecture which was included at the request of several students.}
\end{itemize}
 Lectures 3 to 5 are not included in these proceedings but can be found elsewhere~\cite{Lecture3,Lecture4,Lecture5}. 

The material in these lectures follows loosely  that in my book \cite{LL_book}, together with some 
significant updates (see ref. \cite{LL_book_update}).

\section{Introduction to Lecture 1}
\label{sec:introduction1}

The first lecture, covered in Sections~\ref{sec:introduction1} to~\ref{sec:gaussian}, 
is a recapitulation of material that should already be familiar, but
hopefully with some new emphases. We start with a a discusssion of `What is
Statistics?' and a comparison of `Statistics' and  'Probability'. Next the importance
of calculating uncertainties is emphasised, as well as the difference between random and 
systematic uncertainties. 

The following sections are about combinations. The first is about how to combine different 
individual contributions to a particlar experimental result; the second is the combination of
two or more separate experimental determinaions of the same physical quantity.

The final topics are the Binomial, Poisson and Gaussian probability distributions. Undertanding 
of these is important for many statistical analyses.

\section{What is Statistics?}
Statistics is used to provide quantitative results that give summaries of available data. In High Energy Physics,
there are several different types of statistical activities that are used:
\begin{itemize}
\item{Parameter Determination:\\
We analyse the data in order to extract the best value(s) of one or more parameters in a model. This could be,
for example, the gradient and intercept of a straight line fit to the data; or the mass of the Higgs boson, as deduced using its decay products. In all cases, as well as obtaining the best values of the parameter(s), their uncertainties and possible correlations must be specified. }
\item{Goodness of Fit: \\
We are comparing a single theory with the data, in order to see if they are compatible. If the theory contains free parameters, their best values need to be used to check the Goodness of Fit. If the quality of the fit is unsatisfactory, the best values of the parameters are probably meaningless.    }
\item{Hypothesis Testing: \\   
Here we are comparing the data with two different theories, to see which provides a better description. For example, we may be very interested in knowing whether a model involving the production of a supersymmetric particle is better than one without it. }
\item{Decision Making: \\
As the result of the information we have available, we want to decide what further action to take. For example, we may have some evidence that our data shows hints of an exciting discovery, and need to decide whether we should collect more data. This was the situation faced by the CERN management in 2000, when there were perhaps hints of a Higgs boson in data collected at the LEP Collider.

Such decisions usually require a `cost function' for the various possible outcomes, as well as assessments of their relative probabilities. In the example just quoted, numerical values were needed for the cost of missing an important discovery if the experiment was not continued; and on the other hand of running the LEP Collider for another year and for delaying the 
start of building the Large Hadron Collider.

Decision Making is not considered further in these lectures.     }
\end{itemize}

\section{Probability and Statistics} 
Probability theory involves starting with a model, and using it to make predictions about possible outcomes of an experiment
where randomness plays a role; it involves precise mathematics, and in general there is only one correct solution about
the probabilities of the different outcomes. Statistics involves the opposite procedure of using the observed data in 
order to make statements about the relevant theory or model. This is usually not a precise process and there may be 
different approaches which yield different  answers, none of which being necessarily invalid.

The example of throwing dice (see Table \ref{table:Prob_and_Stats}) illustrates the relationship of Probability 
Theory and Statistics for some of the statistical procedures.

\begin{table} [h !]
\caption{Probability and Statistics: Throwing dice}
\begin{center}
\begin{tabular} {|c|c|c|}
\hline Probability & Statistics &   Procedure \\
\hline Given p(5) =1/6,   &  Given 20 5s in 100 trials, & \\
  what is prob(20 5s in 100 trials)?  &    what is p(5)? &   \\
  & and its uncertainty?  & Parameter Determination \\
\hline If unbiassed,  & Given 60 evens in 100 trials,   &     \\ 
what is prob(n evens in 100 trials)?    & is it unbiassed? & Goodness of Fit\\
\hline     & Or is prob(evens) = 2/3? & Hypothesis Testing \\   
\hline  THEORY $\rightarrow$ DATA  &    DATA $\rightarrow$ THEORY  & \\
\hline
\end{tabular}
\end{center}
\label{table:Prob_and_Stats}
\end{table}

\section{Why uncertainties?}
\label{Uncertainties}
Without an estimate of the uncertainty of a parameter, its central value is essentially useless. This is illustrated by Table 
\ref{table:Harwell_expt}. The three lines of the Table refer to different possible results; all have the same central value of 
the ratio of the experimental result divided by the theoretical prediction, but each has a different uncertainty on this ratio. The 
conclusions about whether the data supports the theory are very different, depending on the magnitude of the uncertainty,
even though the central values are the same for each of the three situations. It is thus crucial to estimate 
uncertainties accurately, and also correlations when measuring two or more parameters.

\begin{table} [h !]
\caption{Experiment testing General Relativity.}
\begin{center}
\begin{tabular} {|c|c|c|}
\hline Experiment/Theory   & Uncertainty   &    Conclusion  \\
\hline 0.970      &  $ \pm$ 0.05     &     Consistent with 1.0 \\
\hline 0.970      &   $\pm$ 0.006  &      Inconsistent  with 1.0 \\
\hline 0.970      &   $\pm$  0.7     &      Do a better experiment \\       
\hline
\end{tabular}
\end{center}
\label{table:Harwell_expt}
\end{table}

\section{Random and systematic uncertainties}
Random or statistical uncertainties result from the limited accuracy of measurements, or from the fluctuations
that arise in counting experiments where the Poisson distribution is relevant (see Section \ref{Poisson}). If the experiment
is repeated, the results will vary somewhat, and the spread of the answers provides (not necessarily the best) 
estimate of the statistical uncertainty. 

Systematic uncertainties can also arise in the measuring process. The quantities we measure may be shifted from the true values.
For example, our measuring device may be miscalibrated, or the number of events we count may be not only from the 
desired signal, but also from various background sources.  Such effects would bias our result, and we should correct for them, for
example by performing some calibration measurement. 
The systematic uncertainty arises from the remaining uncertainty in our corrections.  Systematics can cause a similar shift in a repeated series of experiments, and so, in contrast to statistical uncertainties, they may not be detectable by looking for a spread in 
the results. 

 For example consider a pendulum experiment designed
to measure the acceleration due to gravity $g$ at sea level in a given location:
\begin{equation}
g = 4\pi^2 L/\tau^2
\label{Pendulum}
\end{equation}
where $L$ is the length of the pendulum, $\tau = T/N$ is its period, and $T$ is the time for $N$ oscillations.
The uncertainties we have mentioned so far are the statistical ones on $L$ and 
$T$\footnote{Note that although $N$ involves counting the number of swings, we 
do not have to allow for Poisson fluctuations, since there are no random 
fluctuations involved.}. There may also be systematic uncertainties on these variables.

Unfortunately there are further possible systematics not associated with the measured 
quantities, and which thus require more careful consideration. For example, the 
derivation 
of eqn. (\ref{Pendulum}) assumes that:
\begin{itemize}
\item{our pendulum is simple i.e.  the string is massless, and has a massive bob of infinitesimal 
size;}
\item{the support of the pendulum is rigid;}
\item{the oscillations are of very small amplitude (so that 
$\sin\theta \approx \theta)$; and}
\item{they are undamped.}
\end{itemize}
None of these will  be exact in practice, and so corrections must be estimated for 
them. The uncertainties in these corrections are systematics.

Furthermore, there  may be theoretical uncertainties. For example, we may want the 
value of $g$ at sea level, but the measurements were performed on top of  a mountain. We thus 
need to apply a correction, which depends on our elevation and on the local geology.
There might be two or more different theoretical correction factors, and again this 
will contribute a systematic uncertainty.

\subsection{Presenting the results}
A common way of presenting the result of a measurement $y$ is as  
$y \pm \sigma_{stat} \pm \sigma_{syst}$, where the statistical and 
systematic uncertainties are  
shown separately. Alternatively,  it may be presented as $y\pm \sigma$,
where the total uncertainty is usually given by 
$\sigma^2 = \sigma_{stat}^2 + \sigma_{syst}^2$. 

The other extreme is to give a list of all the individual systematics separately 
(usually in a Table, rather than in the Abstract or Conclusions). The 
motivations for this are that:
\begin {itemize}
\item{systematics are sometimes caused by uncertainties in other people's 
measurements of some relevant quantity. If subsequently this measurement is updated,
it will be possible to reduce the systematic uncertainty appropriately; and}  
\item{our measurement may be combined with others to produce a `World average', 
or it may be used together with another result to calculate something else. 
In both these cases, correlations between the different experimental measurements 
are needed, and so the individual sources are required.}
\end{itemize}

For example, it may be interesting to compare the sea-level values of $g$ at the 
same location several years apart. In that case, although there might be  significant
uncertainties from the correction of the measurements to sea-level, they are 
a fully correlated, and so will cancel in their difference 

\section{Combining uncertainties}
In this section, we consider how to estimate the uncertainty $\sigma_z$ in a quantity of interest $z$, which is defined in terms 
of measured quantities $x, y, ....$  by a known function  $z(x, y,.....)$. The uncertainties on the measured  
quantities are known and assumed to be uncorrelated. The recipe for $\sigma_z$ depends on the functional form of $z$.

\subsection{Linear forms}
As a very simple example, consider
\begin{equation}
z=x-y
\end{equation}
From this, we obtain
\begin{equation}
\delta z = \delta x - \delta y
\label{wrong}
\end{equation}
where $\delta z$ is the change in $z$ that would be produced by specific changes in $x$ and $y$. But  eqn. \ref{wrong}
refers to specific offsets, rather than the uncertainties $\sigma_z$, etc, which are the RMS values of the offsets 
i.e. $ \overline{\delta z^2}$, etc.  Thus we need 
to square eqn.  \ref{wrong}, which yields 
\begin{equation}
\label{squared}
\delta z^2 = \delta x^2 + \delta y^2 -2 \delta x \delta y ,
\end{equation}
and  to average over a whole series of measurements. We then obtain the correct formula for combining the uncertainties:
\begin{equation}
\label{correct}
\sigma_z^2 = \sigma_x^2 + \sigma_y^2,
\end{equation}
provided we ignore the last term in eqn \ref{squared}. The justification for this is that the average value of
$\delta x \delta y$ is zero, provided the uncertainties on $x$ and on $y$ are uncorrelated.\footnote{Note
that it is the {\bf uncertainties} which are required to be uncorrelated. Thus for a simple pendulum, $L$ and $\tau$
are correlated by eqn \ref{Pendulum}, but the uncertainties on the measured length and period are uncorrelated. }

For the general linear form 
\begin{equation}
\label{gen_lin}
z = k_1 x + k_2 y+.......
\end{equation} 
where $k_1, k_2, ....$ are constants, the uncertainty on $z$ is given by
\begin{equation}
\sigma_z = k_1\sigma_ x \  \& \ k_2\sigma_y \  \& \  .......\  ,
\end{equation} 
where the symbol $\&$ is used to mean ``combine using Pythagoras' Theorem". For the special case of $z = x-y$, 
as is expected this gives the result of eqn. \ref{correct} for $\sigma_z$. 

For this case of $z$ being a {\bf linear} function of the measurements, it is the {\bf absolute} uncertainties 
that are relevant for determining $\sigma_z$. It is important {\bf not} to use {\bf fractional} uncertainties. Thus 
if you want to determine your height by making independent measurements of the distances of the top of your 
head and the bottom of your feet from the centre of the earth, each with an accuracy of 1 part in 1000, you will not determine
 your height to anything like 1 part in 1000.


\subsection{Products and quotients}
The general form here is 
\begin{equation}
z = x^\alpha y^\beta ...... , 
\end{equation}
where the powers $\alpha, \beta,$ etc. are constants. This includes  forms such as $x^2,\  y^3/x,  \sqrt{x}/y$, etc.
The formula for combining the uncertainties is
\begin{equation}
\sigma_z /z = \alpha\sigma_x/x\  \&  \ \beta\sigma_y /y \ \&\  .... 
\end{equation}
That is, the {\bf fractional} uncertainty on $z$ is derived from the {\bf fractional} uncertainties on the measurements.

Because this result was derived by taking the first term of a Taylor expansion for $\delta z$, it will be a good approximation only for small uncertainties. If the uncertainties are large, more sophisticated approaches are required for determining the uncertainty in $z$. This also applies to the next section, but is irrelevant for the linear cases discussed above, as all terms in the Taylor series beyond those involving first derivatives are zero.

\subsection{All other functions}
Finally we deal with any functional form $z = z(x_1, x_2, x_3,.....)$. Our prescription of writing down the
first term in the Taylor series expansion for $\delta z$, squaring and averaging gives
\begin{equation}
\sigma_z = \frac{\partial z}{\partial x_1} \sigma_1 \ \& \ \frac{\partial z}{\partial x_2} \sigma_2 \ \& \ ....
\end{equation}
where the $\sigma_i$  are the uncertainties on $x_i$, again assumed uncorrelated.

A slightly easier method to apply is to use a numerical approach for calculating the partial derivatives. We evaluate
\begin{equation}
\begin{split}
z_0  &= z(x_1, x_2, x_3,....)  \\
z_1  &= z(x_1 + \sigma_1, x_2, x_3,....)  \\
z_2  &= z(x_1, x_2 + \sigma_2, x_3,....)  \\
z_3  &= z(x_1, x_2, x_3 + \sigma_3,....)  \\ 
  &\ \ \ \ \ \ \ \ etc.  
\end{split}
\end{equation} 
and then 
\begin{equation} 
\sigma_z^2 = \Sigma (z_i - z_0)^2
\end{equation}

\section{Combining experiments} 
Sometimes different experiments will measure the same physical quantity. It is then reasonable to ask what is our
best information available when these experiments are combined. It is a general rule that it is better to use the
{\bf DATA} for the experiments and then perform a combined analysis, rather than simply combine the {\bf RESULTS}.
However, combining the results is a simpler procedure, and access to the original data is not always possible.

For a series of unbiassed, uncorrelated measurements $x_i$ of the same physical quantity,
the combined value $\hat{x} \pm \hat{\sigma}$ is given by weighting each measurement by $w_i$,
which is proportional to  the inverse of the square of its uncertainty i.e. 
\begin{equation}
\hat{x} = \Sigma w_i x_i, \ \ \ \  w_i =(1/\sigma_i^2)/\Sigma (1/\sigma_j^2) 
\end{equation}
with the uncertainty $\hat{\sigma}$ on the combined value being given by
\begin{equation}
1/\hat{\sigma}^2 = \Sigma 1/\sigma_i^2
\end{equation}
This ensures that the uncertainty on the combination is at least as small as the 
smallest uncertainty of the individual measurements. It should be remembered that the combined uncertainty takes no 
account of whether or not the individual measurements are consistent with each other.
 
In an informal sense, $1/\sigma_i^2$ is the information content of a measurement. Then each $x_i$ is weighted
proportionally to its information content. Also the equation for
$\hat{\sigma^2}$ says that the information content of the combination is the sum of the information contents
of the individual measurements.
 
An example demonstrates that care is needed in applying the formulae. Consider counting the number of high
energy cosmic rays being recorded  by a large counter system for two consecutive one-week periods, with the number of counts being
$100 \pm 10$ and $1 \pm 1$ \footnote{It is vital to be aware that it is a crime (punishable by a forcible transfer
to doing a doctorate on Astrology) to combine such discrepant measurements. It seems likely that someone turned off
the detector between the two runs; or there was a large background in the first measurement which was eliminated
for the second; etc. The only reason for my using such discrepant numbers is to produce a dramatically stupid
result. The effect would have been present with measurements like $100 \pm 10$ and $81 \pm 9$.}.
(See section \ref{Poisson} for the choice of uncertainties). Unthinking application of the formulae
for the combined result give the ridiculous $2 \pm 1$. What has gone wrong?

The answer is that we are supposed to use the {\bf true} accuracies of the individual measurements to assign the weights. 
Here we have used the {\bf estimated} accuracies. Because the estimated uncertainty 
depends on the estimated rate, a downward fluctuation in the measurement results in an underestimated uncertainty,
an overestimated weight, and a downward bias in the combination. In our example, the combination should assume 
that the true rate was the same in the two measurements which used the same detector and which 
lasted the same time as each other, and hence their
true accuracies are (unknown but) equal. So the two measurements should each be given a weight of 0.5, which 
yields the sensible combined result of $50.5  \pm 5$ counts.       

\subsection{\bf BLUE}
A method of combining correlated results is the `{\bf B}est {\bf L}inear {\bf U}nbiassed {\bf E}stimate' ({\bf BLUE}). 
We look for the best linear unbiassed combination
\begin{equation}
x_{BLUE} = \Sigma w_i x_i,
\end{equation} 
where the weights are chosen to give the smallest uncertainty $\sigma_{BLUE}$ on 
$x_{BLUE}$. Also for the combination to be unbiassed, the weights must add up to unity.
They are thus determined by minimising $\Sigma\Sigma w_i w_j E^{-1}_{ij}$, subject to the constraint
$\Sigma w_i = 1$; here $E$ is the covariance matrix for the correlated measurements. 

The $BLUE$ procedure just described is equivalent to the $\chi^2$ approach for checking whether
a correlated set of measurements are consistent with a common value. The advantage of $BLUE$ is that 
it provides the weights for each measurement in the combination. It thus enables us to calculate the contribution 
of various sources of uncertainty in the individual measurements to the uncertainty on the combined result.

\subsection{Why weighted averaging can be better than simple averaging}
Consider a remote island whose inhabitants are very conservative, and no-one leaves or arrives 
except for some anthropologists who wish to determine the number of married people there.
Because the islanders are very traditional, it is necessary to send two teams of anthropologists,
one consisting of males to interview the men, and the other of females for the women. There are too
many islanders to interview them all, so each team interviews a sample and then extrapolates. 
The first team estimates the number of married men as $10,000 \pm 300$. The second, who 
unfortunately have less funding and so can interview only a smaller sample, have a 
larger statistical uncertainty; they estimate $9,000 \pm 900$ married women. Then how many 
married people are there on the island? 

The simple approach is to add the numbers of married men and women, to give $19,000 \pm 950$
married people. But if we use some theoretical input, maybe we can improve the accuracy of 
our estimate. So if we assume that the islanders are monogamous, the numbers of married men and 
women should be equal, as they are both estimates of the number of married couples. The weighted 
average is $9,900 \pm 285$ married couples and hence $19,800 \pm 570$ married people.

The contrast in these results is not so much the difference in the estimates, but that 
incorporating the assumption of monogamy and hence using the weighted average gives a smaller 
uncertainty on the answer. Of course, if our assumption is incorrect, this answer will be biassed.

A Particle Physics example incorporating the same idea of theoretical input reducing the 
uncertainty of a measurement can be found in the `Kinematic Fitting' section of Lecture 2.

\section{Binomial distribution}
This and the next sections on the Poisson and Gaussian distributions are probability theory, in that they make 
statements about the probabilities of different outcomes, assuming that the thoretical distribution is known. However, the
results are important for Statistics, where we use data in order to make statements about theory.

The binomial distribution applies when we have a set of $N$ independent 
trials, in each of which a `success' occurs with probability $p$. Then the 
probability $P(s;N,p)$ of $s$ successes in the $N$ trials is obviously
\begin{equation}
P(s;N,p) = \frac{N!} {s! \ (N-s)!} \  p^s \ (1-p)^{N-s}.
\end{equation}

An example of a Binomial distribution would be the number of times we have a 6 
in 20 throws of a die; or the distribution of the number of successfully reconstructed tracks
in a sample of 100, when the probability  for reconstructing each of them is 0.98

The expected number of successes $<\!\!s\!\!>$ is $\Sigma s\times P(s;N,p)$, which after 
some algebra turns out to be (not surprisingly) $Np$. The variance $\sigma_s^2$ 
of the distribution in $s$ is obviously given by $Np(1-p)$. Note that, while for the Poisson 
distribution the mean and variance are equal, this is not so in general
for the Binomial - it is approximately so at small $p$.

\begin{figure}\begin{center}
\includegraphics[width=0.7\textwidth]{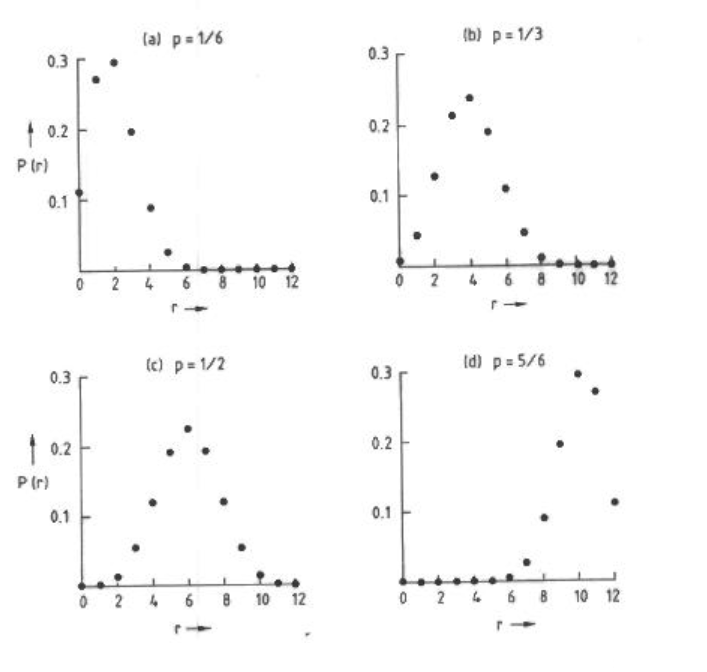}
\caption{The probabilities $P(r)$ according to the binomial distribution, for $r$ successes 
out of 12 independent trials, when the probability $p$ of success in an individual trial is as specified. As the expected number of successes is $12p$, the peak of the distribution moves to the right as $p$ increases. The 
variance of the distribution is $12p(1-p)$ and hence is largest for $p=1/2$. Since the chance of 
success when $p=1/6$ is the same as that for failure when $p=5/6$, diagrams (a) and (d) are mirror images of each other. Similarly for $p=1/2$ (see (c)) the distribution is symmetric about $r=6$ successes. }   
\label{fig:Binomials}
\end{center}
\end{figure}

As an example several Binomial distributions with fixed number of 
trials $N$ but varying probabilities of success $p$ are shown in Fig.~\ref{fig:Binomials}.

\section{Poisson distribution}
\label{Poisson}

\begin{figure}
\begin{center}
\includegraphics[width=0.37\textwidth]{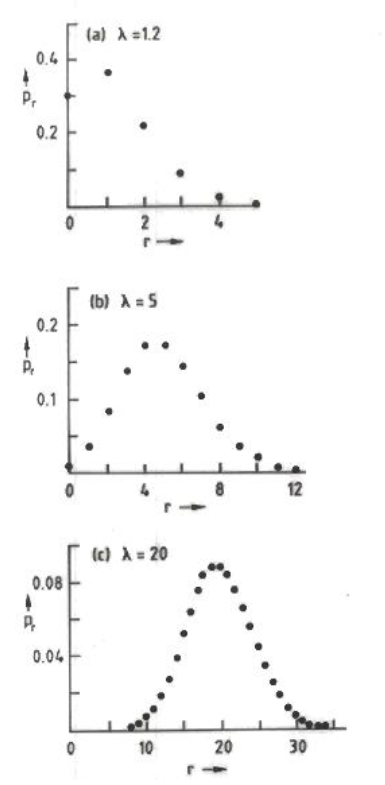}
\caption{ Poisson distributions for various values of the Poisson parameter $\lambda$: (a) $\lambda = 1.2$  (b) $\lambda = 5.0$
(c) $\lambda = 20.0$. $P_r$ is the probability for observing $r$ events. For each $\lambda$, the mean value of $r$ is 
$\lambda$ and the RMS width is $\sqrt\lambda$. As $\lambda$ increases above about 10, the distribution becomes more like
a Gaussian. }   
\label{fig:Poissons}
\end{center}
\end{figure}

The Poisson distribution (see Fig.~\ref{fig:Poissons}) applies to situations where we are counting a series of observations which are occuring randomly and independently during a fixed time interval $t$, where the underlying rate $r$ is constant. The observed number $n$ will fluctuate when the experiment is repeated, and can in principle take any integer value from zero to infinity. The Poisson probabilty of observing $n$ decays is given by
\begin{equation}
P_n = e^{-rt} (rt)^n/n!
\end{equation}
It applies to the number of decays observed from a large number $N$ of radioactive nuclei, when the observation time $t$ is small compared to the lifetime $\tau$.  It will not apply if  $t$ is much larger than $\tau$, or if the detection system has a dead time, so that after observing a decay the detector cannot observe another decay for a period $T_{dead}$.

Another example is the number of counts in any specific bin of a histogram when the data is accumulated over a fixed time.  

The average number of observations is given by 
\begin{equation}
<n> = \Sigma n P_n = rt
\end{equation}
If we write the expected number as $\mu$, the Poisson probability becomes
\begin{equation}
 P_n = e^{-\mu} \mu^n/n!
\end{equation}
It is also relatively easy to show that the variance 
\begin{equation}
\sigma^2 = \Sigma (n - \mu)^2 P_n = \mu
\end{equation}
This leads to the well-known $n \pm \sqrt n$ approximation for the value of the Poisson parameter when we have $n$ counts. This approximation is, however, particularly bad when there are zero observed events; then $0\pm0$ incorrectly suggests that the Poisson parameter can be only zero.

Poisson probabilities can be regarded as the limit of Binomial ones as the number of trials $N$ tends to infinity and the Binomial probability of success $p$ tends to zero, but the product $Np$ remains constant at $\mu$.  

When the Poisson mean becomes large, the distribution of observed counts approximates to a Gaussian (although the Gaussian is a continuous distribution extending down to $-\infty$, while a Poisson observable  can only take on non-negative integral values). This approximation is useful for the $\chi^2$ method for parameter estimation and goodness of fit (see Lecture 2).

\subsection{Relation of Poisson and Binomial Distributions}
An interesting example of the relationship between the Poisson and Binomial distributions is exhibited by the following example.

Imagine that the number of people attending a series of lectures is Poisson distributed with a constant 
mean $\nu$, and that the fraction of them who are male is $p$. Then the overall probability $P$ of 
having N people of  whom $M$ are male and $F = N - M$ are female is given by the product of the Poisson 
probability $P_{pois}$ for $N$ and the binomial probability $P_{bin}$ for $M$ of the $N$ people being male. i.e.
\begin{equation}
P = P_{pois} P_{bin} = \frac{e^{-\nu} \nu^N}{N!} \times \frac{N!}{M!F!} p^M (1-p)^F
\end{equation} 
This can be rearranged as
\begin{equation}
P =  \frac{e^{-\nu p} (\nu p)^M}{M!} \times \frac{e^{-\nu (1-p)} (\nu(1-p))^F}{F!}
\end{equation} 
This is the product of two Poissons, one with Poisson parameter $\nu p$, the expected number of males, and the other with parameter $\nu(1-p)$, the expected number of females. Thus with a Poisson-varying total number of observations,  divided into two  categories (here male and female), we can regard this as Poissonian in the total number and Binomial in the  separate categories, or as two independent Poissons, one for each category. Other situations to which this applies could be radioactive nuclei, with decays detected in the forward or backward hemispheres; cosmic ray showers, initiated by protons or by heavier nuclei; patients arriving at a hospital emergency centre, who survive or who die; etc.

\subsection{For your thought}
The first few Poisson probabilities $P(n;\mu)$ are
\begin{equation} 
\begin{split}
P(0) = e^{-\mu}, \ \ \ \ \ P(1) = \mu e^{-\mu}, \ \ \ \ \ P(2) = (\mu^2/2!)\ e^{-\mu}, \ \ \ \ \ etc.
\end{split}
\end{equation}
Thus for small $\mu$,  $P(1)$ and $P(2)$ are approximately $\mu$ and $\mu^2/2$ respectively. But if the probability
of one rare event happening is $\mu$, why is the probability for 2 independent rare events not equal to $\mu^2$?

\section{Gaussian distribution}
\label{sec:gaussian}

\begin{figure}
\begin{center}
\includegraphics[width=0.57\textwidth]{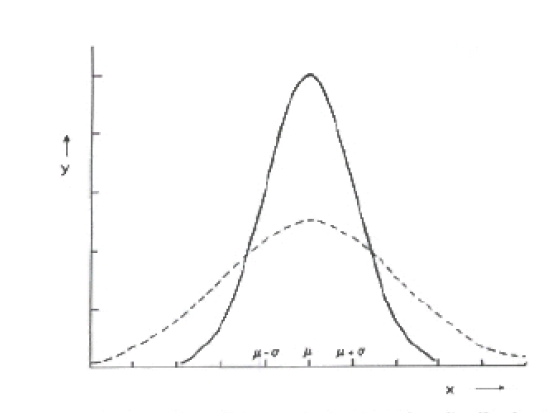}
\caption{Gaussian distributions. Both are centred at $x = \mu$, but the dashed curve is twice as wide as the solid one. Because
they have the same normalisation the maximum of the solid curve curve is twice as high as that of the dashed one. 
The scale on the horizontal axis refers to the solid curve.  }   
\label{fig:Gaussians}
\end{center}
\end{figure}

The Gaussian or normal distribution (shown in Fig.~\ref{fig:Gaussians}) is of widespread usage in data analysis. Under suitable conditions, in
a repeated series of measurements $x$ with accuracy $\sigma$ when the true value of the quantity
is $\mu$, the distribution of $x$  is given by a Gaussian\footnote{However, it is often the case 
that such a distribution has heavier tails than the Gaussian.}. A mathematical motivation is given by the 
Central Limit Theorem, which states that the sum of
a large number of variables with (almost) any distributions is approximately Gaussian.   

For the Gaussian, the probability density $y(x)$  of an observation $x$  is given by
\begin{equation}
y(x) = \frac{1}{\sqrt{2\pi} \sigma} e^{-\frac{(x-\mu)^2}{2\sigma^2}}
\end{equation}
where the parameters $\mu$ and $\sigma$ are respectively the centre and width of the distribution.
The factor 
$1/(\sqrt{2\pi} \sigma)$ is required to normalise the area under the curve, so that $y(x)$
can be directly interpreted as a probability density.

There are several properties of $\sigma$:
\begin{itemize}
\item{The mean value of $x$ is $\mu$, and the standard deviation of its distribution is $\sigma$. Since the usual symbol for  
standard deviation is $\sigma$, this leads to the formula $\sigma = \sigma$ (which is not so trivial as it seems, since the 
two $\sigma$s have different meanings). This explains the curious factor of 2 in the denominator of the exponential,
since without it, the two types of $\sigma$ would not be equal.}
\item{The value of $y$ at the $\mu \pm \sigma$ is equal to the peak height multiplied  by  $e^{-0.5}$ = 0.61.
If we are prepared to overlook the difference between 0.61 and 0.5, $\sigma$ is the half-width of the distribution at
`half' the peak height.}
\item{The fractional area in the range $x = \mu - \sigma$ to $\mu + \sigma$ is 0.68. Thus for a series of unbiassed, independent Gaussian 
distributed measurements}, about 2/3 are expected to lie within $\sigma$ of the true value.
\item{The peak height of $y$ at $x=\mu$  is 
$1/(\sqrt{2\pi} \sigma)$. It is reasonable that this is proportional to 
$1/\sigma$ as the width is proportional to $\sigma$, so $\sigma$ cancels out in the product
of the height and width, as is 
required for a distribution normalised to unity.}
\end{itemize}

\begin{figure}
\begin{center}
\includegraphics[width=0.7\textwidth]{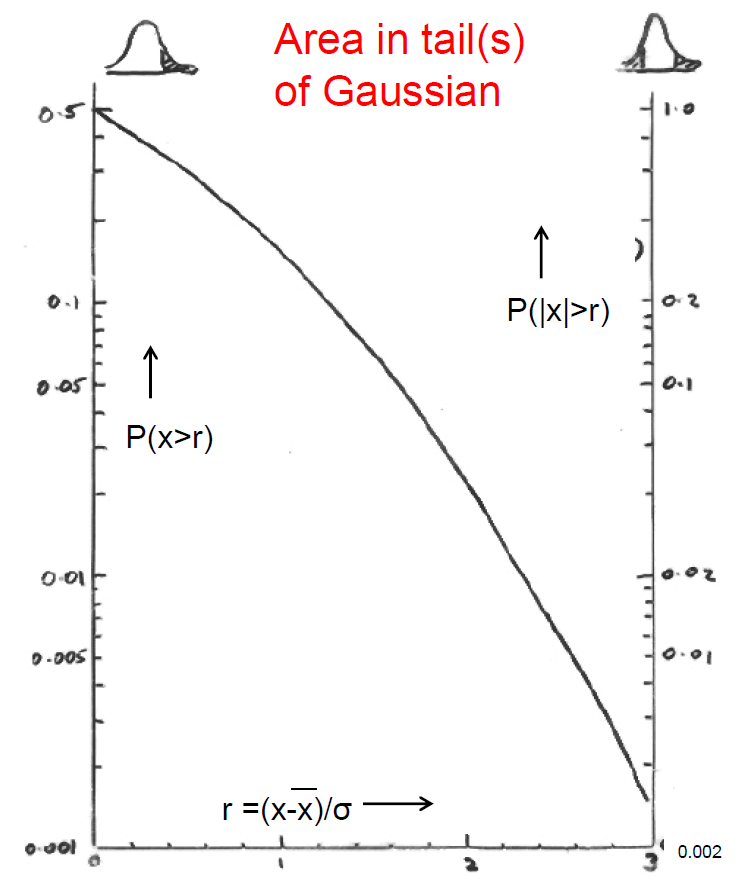}
\caption{The fractional area in the tail(s) of a Gaussian distribution i.e. the area with $f$ above some specified value
$r$,
where $f$ is the distance from the mean, measured in units of the standard deviation $\sigma$. The scale on the 
left refers to the one-sided tail, while that on the right is for both tails. Thus for $r = 0$, the 
fractional areas are 1/2 and 1 respectively. }   
\label{fig:Gauss_tail}
\end{center}
\end{figure}

For deciding whether an experimental measurement is consistent with a theory, more useful  than the Gaussian distribution 
itself is its tail area beyond $r$, a number of standard deviations from the central value (see Fig.~\ref{fig:Gauss_tail}). 
This gives the  probability of obtaining a result as extreme as ours or more so as a consequence of statistical fluctuations, 
assuming that the theory is correct (and that our measurement is unbiassed, it is Gaussian distributed, etc.). If this 
probability is small, the measurement and the theory may be inconsistent. 

Figure~\ref{fig:Gauss_tail} has two different vertical scales, the left one for the probability of a fluctuation in a specific 
direction, and the right side for a fluctuation in either direction. Which to use depends on the particular
situation. For example if we were performing a neutrino oscillation disappearance experiment, we would be looking for 
a reduction in the number of events as compared with the no-oscillation scenario, and hence would be interested in 
just the single-sided tail. In contrast searching for any deviation from the Standard Model expectation, maybe the two-sided tails would be more relevant.   








\section{Introduction to Lecture 2}
\label{sec:introduction2}
This lecture deals with two different methods for determining parameters, least squares and likelihood, 
when a functional form is fitted
to our data. A simple example would be straight line fitting, where the parameters are the intercept and gradient
of the line. However the methods are much more general than this. Also there are other
methods of extracting parameters; these include the more fundamental Bayesian and Frequentist methods,
which are dealt with in Lecture 3 . 

The least squares method also provides a measure of Goodness of Fit for the agreement between the theory with the 
best values of the parameters, and the data; this is dealt with in section \ref{GofF}. The likelihood technique plays 
an important role in the Bayes approach, and likelihood ratios are relevant for choosing between two hypotheses;
this is covered in Lecture 4. 

\section{Least squares: Basic idea}
As a specific example, we will consider fitting a straight line $y = a + bx$ to some data, which consist of a series on $n$ data
points, each of which specifies $(x_i, y_i \pm \sigma_i)$ i.e. at precisely known $x_i$, the $y$ co-ordinate is measured
with an uncertainty $\sigma_i$. The $\sigma_i$ are assumed to be uncorrelated. The more general case could involve  
\begin{itemize}
\item{a more complicated functional form than linear;}
\item{multidimensional $x$  and/or $y$;}
\item{correlations among the $\sigma_i$; and}
\item{uncertainties on the $x_i$ values.}
\end{itemize} 

In Particle Physics, we often deal with a histogram of some physical quantity $x$ (e.g. mass, angle, 
transverse momentum, etc.), in which case $y$ is simply the number of counts for that $x$ bin. Another possiblity
is that $y$ and $x$ are both physical quantities e.g. we have a two-dimensional plot showing the recession velocities 
of galaxies as a function their distance.    

There are two statistical issues: Are our data consistent with the theory i.e. a straight line? And what are 
the best estimates of the parameters, the intercept and the gradient? The former is a Goodness of Fit 
issue, while the latter is Parameter Determination. The Goodness of Fit is more fundamental, in that
if the data are not consistent with the hypothesis, the parameter values are meaningless. However, we will first
consider Parameter Detemination, since checking the quality of the fit requires us to use the best straight
line. 

The data statistic used for both questions is $S$, the weighted sum of squared discrepancies\footnote{Many people 
refer to this as $\chi^2$. I prefer S, because otherwise a discussion about whether or not $\chi^2$ follows the
mathematical $\chi^2$ distribution sounds confusing.}
\begin{equation}
S = \Sigma (y_i^{th} - y_i^{obs})^2/\sigma_i^2 = \Sigma (a + bx_i - y_i^{obs})^2 / \sigma_i^2
\label{eqn:S}
\end{equation}
where $y_i^{th} = a + bx_i$ is the predicted value of $y$ at $x_i$, and $y_i^{obs}$ is the observed value.  In the 
expression for $S$, we regard the data $(x_i, y_i \pm \sigma_i)$ as being fixed, and the parameters $a$
and $b$ as being variable.
If for specific values of $a$ and $b$ the predicted values of $y$ and the corresponding observed ones are all close 
(as measured in terms of the 
uncertainties $\sigma$), then $S$ will be `small', while significant discrepancies result in large $S$. Thus, according
to the least squares method, the best values of the parameters are those that minimise $S$, and the width of the $S$
distribution determines their uncertainties. For a good fit, the value of $S_{min}$ should be `small'. A more quantative
discussion of `small' appears below. 

To determine the best values of $a$ and $b$, we need to set the first  derivatives of $S$  with respect to $a$ and $b$ both
equal to zero. This leads to two simultaneous linear equations for $a$ and $b$ \footnote{The derivatives are linear in the 
parameters, because the functional form is linear in them. This would also be true for more complicated situations 
such as a higher order polynomial (Yes, with respect to the coefficients, a $10^{th}$ order polynomial is linear),
a series of inverse powers, Fourier series, etc.}  
which are readily solved, to yield
\begin{equation}
\begin{split}
a&=\frac{<x^2><y> - <xy><x>}{<x^2> - <x>^2} \\
b&=\frac {<xy> - <x> <y>}{<x^2> - <x>^2} 
\end{split}
\end{equation} 
where $<f>\ = \Sigma (f_i/\sigma_i^2) /  \Sigma (1/\sigma_i^2)$  i.e it is the weighted average of the quantity inside the
brackets. If the positions of the data points are such that $<x>\ = 0$, then  $a=\ <y>$, i.e. the height of the
 best fit line at the weighted centre of gravity of the data points is just the weighted average of the $y$ values.

It is also essential to calculate the uncertainties $\sigma_a$ and $\sigma_b$ on the parameters and their correlation
coefficient $\rho = cov/(\sigma_x \sigma_y)$, where $cov$ is their covariance. The elements of the inverse
covariance matrix $M$ are given by
\begin{equation}
\begin{split}
M_{aa} &=\frac{1}{2} \frac{\partial^2S}{\partial a^2} = \Sigma (1/\sigma_i^2)   \\
M_{ab} &=\frac{1}{2} \frac{\partial^2S}{\partial a \ \partial b} =   \Sigma(x_i/\sigma_i^2) \\
M_{bb} &=\frac{1}{2} \frac{\partial^2S}{\partial b^2} = \Sigma(x_i^2/\sigma_i^2) \\
\end{split}
\end{equation}
The covariance matrix is obtained by inverting $M$. Since the covariance is proportional to $-< x >$, if the 
data are centred around $x = 0$, the uncertainties on $a$ and $b$ will be uncorrelated. That is 
 one reason why track parameters are usually specified at the centre of the track, rather than at its starting point.

\subsection{Correlated uncertainties on data}
So far we have considered that the uncertainties on the data are uncorrelated, but this is not always the case; correlations can 
arise from some common systematic. Then instead of the first equation of  (\ref{eqn:S}), we use
\begin{equation}
S = \Sigma\Sigma (y_i^{th} - y_i^{obs})E_{ij} (y_j^{th} - y_j^{obs})
\label{correlated_S}
\end{equation}
where the double  summation is over $i$ and $j$, and $E$ is the inverse covariance matrix\footnote{We use the symbol $E$ for the inverse covariance matrix of the measured variables $y$, and $M$ for that of the output parameters (e.g. $a$ and $b$
for the straight line fit).}
for the uncertainties on the 
$y_i$. For the special case of uncorrelated  uncertainties, the only non-zero elements of $E$ are the diagonal ones 
$E_{ii} = 1/\sigma_i^2$ and then  
eqn. (\ref{correlated_S}) reduces to (\ref{eqn:S}).

This new equation for $S$ can then be minimised to give the best values of the parameters, and $S_{min}$ can be used in a Goodness of Fit test. As before, if $y^{th}$ is linear in the parameters, their best estimates can be obtained by solving simultaneous linear equations, without the need for a minimisation programme.

\section{Least squares for Goodness of Fit}
 \label{GofF}
\subsection{The chi-squared distribution}
It turns out that, if we repeated our experiment a large number of times, and certain conditions are satisfied, 
then $S_{min}$ will follow a $\chi^2$ distribution with $\nu = n - p$ degrees of freedom, where $n$ is the 
number of data points, $p$ is the number of free parameters in the fit, and $S_{min}$ is the value of $S$ 
for the best values of the free parameters. For example, a straight line with free intercept and gradient fitted 
to 12 data points would have $\nu = 10$.

The conditions for this to be true include:
\begin{itemize}
\item{the theory is correct:}
\item{the data are unbiassed and asymptotic;}
\item{the $y_i$ are Gaussian distributed about their true values;}
\item{the estimates for $\sigma_i$ are correct;$\ \ \ \ \ $ etc. }
\end{itemize}

Useful properties to know about the mathematical $\chi^2$ distribution are that their mean is $\nu$ and their variance is
$2\nu$. Thus if a global fit to a lot of data has $S_{min}$ = 2200 and there are 2000 degrees of freedom, we can 
immediately estimate that this is equivalent to a fluctuation of 3.2$\sigma$.

\begin{figure}
\begin{center}
\includegraphics[width=0.6\textwidth]{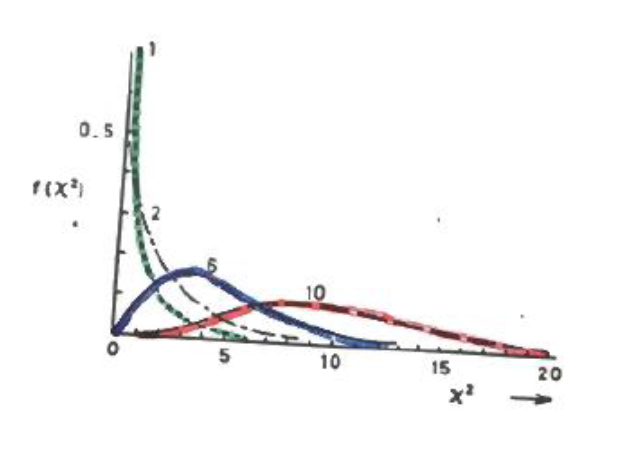}
\caption{
Mathematical distributions of $\chi^2$, for different numbers of degrees of freedom $\nu$ (shown beside each curve).
As $\nu$ increases, so do the mean and variance of the distribution. }  
\label{fig:chi_squared}
\end{center}
\end{figure}

\begin{figure}
\begin{center}
\includegraphics[width=0.75\textwidth]{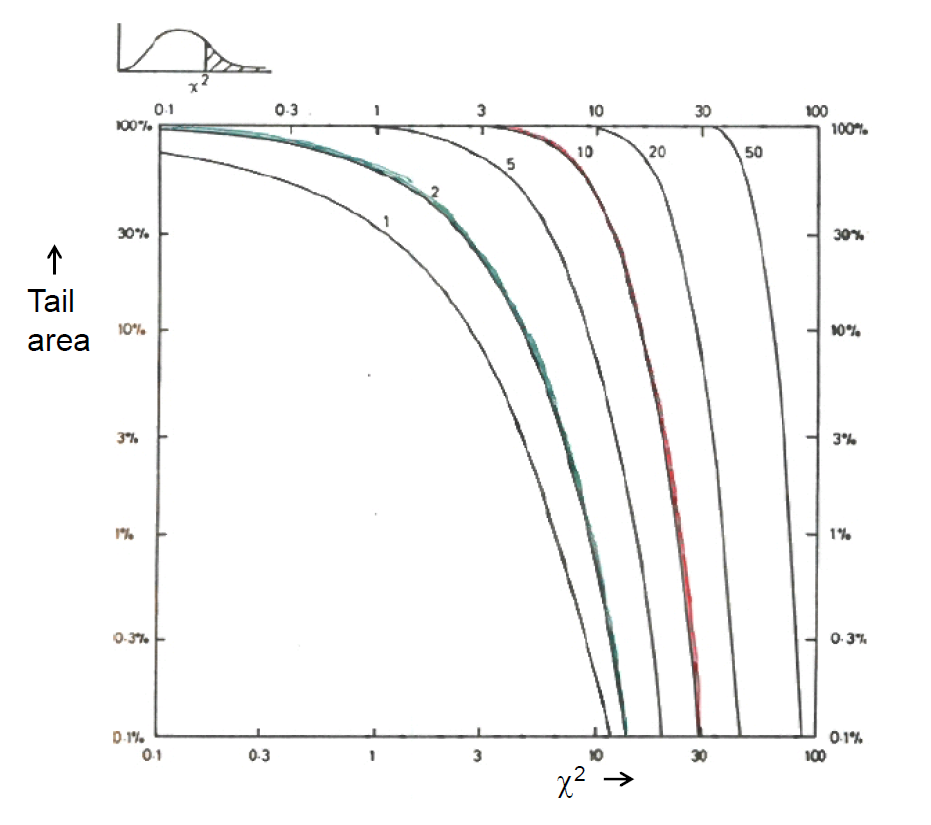}
\caption{ The percentage area in the upper tails of $\chi^2$ distributions, for various numbers of degrees of freedom, 
shown by each curve. Both scales are logarithmic. These curves bear the same relationship to those of
figure \ref{fig:chi_squared} as does fig. 4 to the Gaussian of fig. 3, both in Lecture 1. }  
\label{fig:chi_sq_tail}
\end{center}
\end{figure}

More useful than plots of $\chi^2$ distributions are those of the fractional tail area beyond a particular value 
of  $\chi^2$ (see figs. \ref{fig:chi_squared} and \ref{fig:chi_sq_tail} respectively).
The $\chi^2$ goodness of fit test consists of
\begin{itemize}
\item{For the given theoretical form, find the best values of its free parameters, and hence $S_{min}$;}
\item{Determine $\nu$ from $n$ and $p$; and}
\item{Use $S_{min}$ and $\nu$ to obtain the tail probability $p$ \footnote{If the conditions for $S_{min}$ to follow a 
$\chi^2$ distribution are satisfied, this simply involves using the tail probability of a $\chi^2$ distribution. 
In other cases, it may be necessary to use Monte Carlo simulation to obtain the distribution of $S_{min}$;
this could be tedious.  }.} 
\end{itemize}

Then $p$ is the probability that, if the theory is correct, by random fluctuations we would have obtained a value of $S_{min}$ at least as large as the observed one. If this probability is smaller than some pre-defined level $\alpha$, we reject the hypothesis that the model provides a good description of the data.

\subsection{When $\nu \ne n - p$}
If we add an extra  parameter into our theoretical description, even if it is not really needed, we expect the value of $S_{min}$ to decrease slightly. (This contrasts with including a parameter which is really relevant, which can result in a dramatic reduction in $S_{min}$.)  In determining $p$-values, this is allowed for by the reduction of $\nu$. On average,
a parameter which is not needed reduces $S_{min}$ by 1. But consider the following examples.
\subsubsection{Small oscillatory term}
Imaging we are fitting a histogram of a variable $\phi$ by a distribution of the form
\begin{equation}
\frac{dy}{d\phi} = N[ 1 + 10^{-6} cos(\phi -\phi_0)],
\end{equation}
where the two parameters are the normaisation $N$ and the phase $\phi_0$. Because of the factor $10^{-6}$ in front
of the cosine term,  $\phi_0$ will have a miniscule effect on the prediction, and so including this as a parameter has negligible effect on $S_{min}$; $\phi_0$ is effectively not a free parameter.

\subsubsection{Neutrino oscillations} 
For a scenario of two oscillating neutrino flavours, the probability $P$ of a neutrino of energy $E$  to remain the same flavour after
a flight length $L$  is
\begin{equation}
P = 1 - A sin^2(\delta m^2 L/E)
\label{neutrino}
\end{equation}
where the two parameters are $\delta m^2$, the difference in the mass-squareds of the two neutrino flavours, and 
$A = sin^2 2\theta$ with $\theta$ being the mixing angle. However, since for small angles $\alpha,\ sin\alpha\approx \alpha$, for small $\delta m^2L/E$ the probability $P$ of eqn \ref{neutrino} is approximately $1 - A(\delta m^2 L/E)^2$. Thus the two parameters occur only as the product  $A (\delta m^2)^2$, and cannot be determined separately. Thus  in that regime we have effectively just a single parameter.
\vspace{0.2in}

In both the above examples, an enormous amount of data would enable us to distinguish the small effects produced by the second 
parameter; hence the requirement for asymptotic conditions. 

\subsection{Errors of First and Second Kind}
In deciding in a Goodness of Fit test whether or not to reject the null  hypothesis  $H_0$ (e.g. that the data points lie on a 
straight line), there are two sorts of mistake we might make:
\begin{itemize}
\item{Error of the First Kind. This is when we reject $H_0$ when it is in fact true. The fraction of cases in which this happens  
should equal  $\alpha$, the cut on the $p$-value. }
\item{Error of the Second Kind. This is when we do not reject $H_0$, even though some other hypothesis is true. The rate at which this happens depends on how similar $H_0$ and the alternative hypothesis are, the relative frequencies of the two hypotheses being true, etc.}
\end{itemize}
As $\alpha$ increases the rates of Errors of the First and Second kinds go up and down respectively. 
These Errors correspond to a loss of efficiency and to an increase of contamination respectively.



\subsection{Other Goodness of Fit tests}
The $\chi^2$ method is by no means the only one for testing Goodness of Fit.
Indeed whole books have been written on the subject\cite{DAgostino}. Here
we mention just one other, the Kolmogorov-Smirnov method (K-S), which has the 
advantage of working with individual observations. It thus can be used with 
fewer observations than are required for the binned histograms in the $\chi^2$
approach. 

\begin{figure}
\begin{center}
\includegraphics[width=0.5\textwidth]{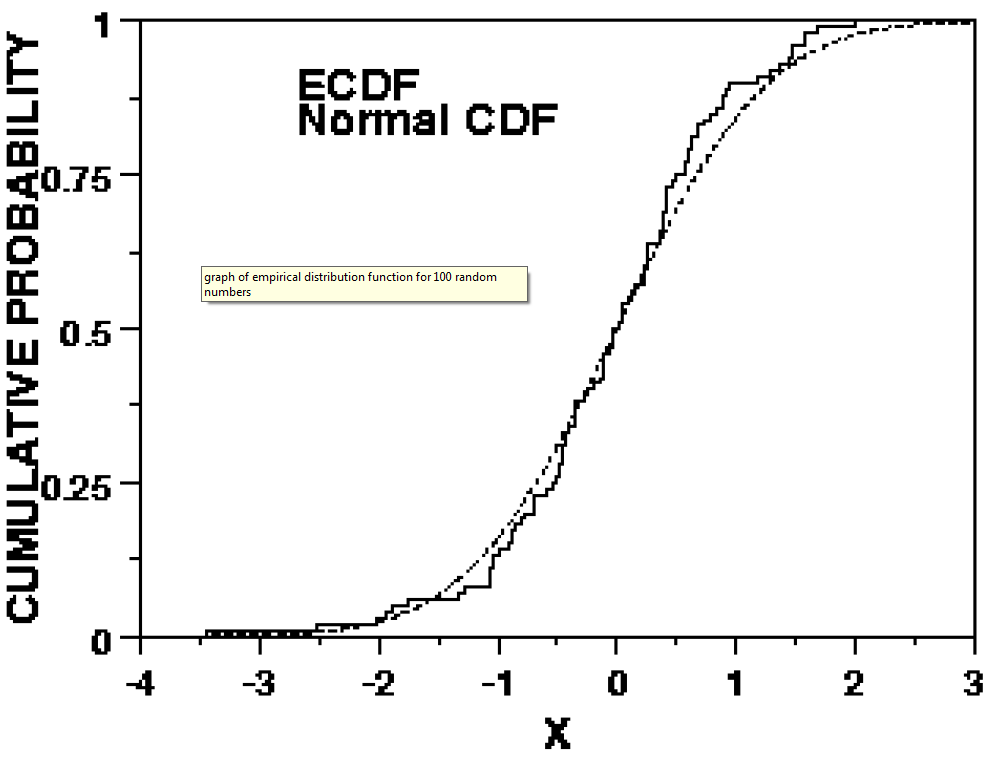}
\caption{Cumulative distributions for the Kolmogorov-Smirnov goodness of fit method. The stepped distribution shows the 
fraction of events in a data sample, while the continuous curve is that expected for a Gaussian with mean zero 
and unit variance. The method uses the maximum vertical separation $d$ between the two distributions, and the number
of observations, to obtain the probability of obtaining a value of $d$ at least as large as the observed one.
A small probability implies that it is unlikely that the data sample comes from the assumed distribution. }  
\label{fig:K_S}
\end{center}
\end{figure}

A cumulative plot is produced of the fraction of events as a function of the variable
of interest $x$. An example is shown in Fig.~\ref{fig:K_S}. 
This shows the fraction of data events with $x$ smaller than any particular 
value. It is thus a stepped plot, with the fraction going from zero at the extreme left, 
to unity on the right hand side. Also on the plot is a curve showing the expected cumulative
fraction for some theory. The K-S method makes use of the largest (as a function of $x$) vertical 
discrepancy $d$ between the data plot and the theoretical curve. Assuming the theory is true
and given the number of observations $N$, the probability $p_{KS}$ of obtaining $d$ at least as large 
as the observed value can be calculated. The beauty of the K-S method is that this probability
is independent of the details of the theory. As in the $\chi^2$ approach, the K-S probability 
gives a numerical way of checking the compatibility of theory and data. If $p_{KS}$ is small, we 
are likely to reject the theory as being a good description of the data.

Some features of the K-S method are:
\begin{itemize}
\item{The main advantage is that it can use a small number of observations.}
\item{The calculation of the K-S probability depends on there being no adjustable parameters in the theory.
If there are, it will be necessary for you to determine the expected distribution for $d$, presumably 
by Monte Carlo simulation.}
\item{It does not extend naturally to data of more than one dimension, because of there being no unique
way of producing an ordering in several dimensions.}
\item{It is not very sensitive to deviations in the tails of distributions, which is where searches for
new physics are often concentrated e.g. high mass or transverse momentum. Fortunately variants of K-S exist, 
which put more emphasis on discrepancies in the tails.}
\item{Instead of comparing a data cumulative distribution with a theoretical curve, it can alternatively be
compared with another distribution. This can be from a simulation of a theory, or with another data set. The
latter could be to check that two data sets are compatible. 
The calculation of the K-S probability now requires the maximum discrepancy $d$, and the numbers of events 
$N_1$ and $N_2$ in each of the two distributions being compared.}
\end{itemize}

\section{Kinematic Fitting}
Earlier we had the example of estimating the number of married people on 
an island, and saw that introducing theoretical information could improve
the accuracy of our answer. Here we use the same idea in the context of
estimating the momenta and directions of objects produced in a high energy
interaction. The theory we use is that energy and momentum are conserved
between the inital state collison and the observed objects in the reaction.

The reaction can be either at a collider or with a stationary target. We 
denote it by $a + b \rightarrow c + d + e$, but the number of final state 
objects can be arbitrary. We assume for the time being the energy and 
momenta of all the objects are measured\footnote{For objects like charged 
particles whose momenta are determined from their trajectories in a 
magnetic field, the energy is determined from the momentum by using the 
relevant particle mass.}.

The technique is to consider all possible configurations of the particles'
kinematic variables that conserve momentum and energy, and to choose that 
configuration that is closest to the measured variables. The degree of
closeness is defined by the weighted sum of squares of the discrepancies $S$,
taking the uncertainties and correlations into account. If the uncertainties 
on the kinematic quantities $m_i$ were uncorrelated,   
\begin{equation}
S = \Sigma (f_i - m_i)^2/\sigma_i^2
\end{equation}
where the summation is over the 4 components for all the objects in the 
reaction, $m_i$ are the measured values and $f_i$ are the corresponding 
fitting quantities. Because of correlations, however, this becomes
\begin{equation}
S = \Sigma\Sigma (f_i - m_i) E_{ij} (f_j - m_j) 
\end{equation}
where there is now a double summation over the components, and $E_{ij}$ is the 
$(i,j)^{th}$ component of the inverse covariance matrix for the measured 
quantities\footnote{The main correlations are among the 4 components of a single
object, rather than between different objects.}. 
The procedure then consists in varying $f$ in order to minimise $S$, subject to
the energy and momentum constraints. This usually involves Lagrange Multipliers.
The result of this procedure is to produce a set of fitted values of all the 
kinematic quantities, which will have smaller uncertainties than the measured ones.
This is an example of incorporating theory to improve the results. Thus if the objects 
are jets, their directions are usually quite well determined, but their energies less so. 
The fitting procedure enables the accurately determined jet directions to help 
reduce the uncertainties on the jet energies.

The fitting procedure 
also provides $S_{min}$, which is a measure of how well the best $f_i$ agree with the $m_i$.
In the case described, the distribution of $S_{min}$ is approximately $\chi^2$ 
with 4 degrees of freedom (because of the 4 constraints).

If $S_{min}$ is too large, then our assumed hypothesis for the reaction may be 
incorrect; for example, there might have been an extra object produced in the 
collision that was undetected (e.g. a neutrino, or a charged particle which passed 
through an uninstrumented region of our detector).

Since we have 4 constraint equations, we can also allow for up to 4 missing kinematic
quantities. Examples include an undetected neutrino in the final state (3 unmeasured 
momentum components), a wide-band neutrino beam of known direction (1 missing variable),
etc. With $m$ missing variables in an interaction involving a single vertex, 
$S_{min}$ should have a $\chi^2$ distribution with $4-m$ degrees of freedom.

Kinematic fitting can be extended to more complicated event topologies including production
and decay vertices, reactions involving particles of well known mass which decay 
promptly (e.g. $\psi \rightarrow \mu^+ \mu^-$), etc.

\subsection{Example of a simplified kinematic fit}
Consider a non-relativistic elastic scattering of two equal mass objects, for example a slow 
anti-proton hitting a stationary proton. 
For simplicity, the measured angles $\theta_1^m \pm \sigma$ and $\theta_2^m \pm \sigma$ that 
the outgoing particles make with the direction
of travel of the incident anti-proton are assumed to have the same uncorrelated 
uncertainties $\sigma$. As a result of energy and momentum conservation, the angles must
satisfy the constraint
\begin{equation}
\label{constraint}
\theta_1^t + \theta_2^t = \pi/2
\end{equation}
where the superscipt $t$ denotes the true value. There are 3 further constraints 
but for simplicity we shall ignore them.

To find our best estimates of $\theta_1^t$ and $\theta_2^t$, we must minimise
\begin{equation}
S = (\theta_1^t - \theta_1^m)^2/\sigma^2 + (\theta_2^t - \theta_2^m)^2/\sigma^2
\end{equation}
subject to the constraint \ref{constraint}. By using Lagrange Multipliers or by 
eliminating $\theta_2^t$ and then minimising $S$, this yields
\begin{equation}
\begin{split}
\label{fitted}
\theta_1^t &= \theta_1^m + 0.5*(\pi/2 - \theta_1^m - \theta_2^m) \\ 
\theta_2^t &= \theta_2^m + 0.5*(\pi/2 - \theta_1^m - \theta_2^m)
\end{split}
\end{equation}
That is, the best estimate of each true value is obtained by adding to the corresponding measured value
half the amount by which the measured values fail to satisfy the constraint 
\ref{constraint}. 

The uncertainties on the fitted estimates of the angles are easily obtained by 
propagation of the uncertainties $\sigma$ on the measured angles vias eqns. 
\ref{fitted}, and are both equal to $\sigma/\sqrt 2$. 

We thus have an example 
of the promised outcome that kinematic fitting improves the accuracy of our 
measurements. The factor of $\sqrt 2$ improvement can easily be understood in that 
we have two independent estimates of $\theta_1^t$, the first being the 
original measurement $\theta_1^m$, and the other coming from the measurement 
$\theta_2^m$ via the constraint \ref{constraint}.  However, even with uncorrelated uncertainties
on the measured angles, the fitted ones would be anti-correlated.

\section{THE paradox}
I refer to this as `THE' paradox as, in various forms, it is the basis of the
most frequently asked question.

You have a histogram of 100 bins containing some data, and use this to determine the 
best value $\mu_0$ of a parameter $\mu$ by the $\chi^2$ method. It turns out that $S_{min} =87$,
which is reasonable as the expected value for a $\chi^2$ with 99 degrees of freedom is $99\pm14$.
A theorist asks whether his predicted value $\mu_{th}$  is consistent with your data,
so you calculate $S(\mu_{th})$ = 112. The theorist is happy because this is within 
the expected range. But you point out that the uncertainty in $\mu$ is calculated by finding where
$S$ increases by 1 unit from its minimum. Since 112 is 25 units larger than 87, this is equivalent 
to a 5 standard deviation discrepancy, and so you rule out the theorist's value of $\mu$.

Deciding which viewpoint is correct is left as an excercise for the reader.

\section{Likelihood}
The likelihood function is very widely used in many statistics applications. In this 
Section, we consider it just for Parameter Determination. An important feature of the 
likelihood approach is that it can be used with {\bf unbinned} data, and
hence can be applied in situations where there are not enough individual observations
to construct a histogram for the $\chi^2$ approach. 

We start by assuming that we wish to fit our data $x$, using a model $f(x;\mu)$ 
which has one or more free parameters $\mu$, whose value(s) we need to determine. 
The function $f$ is known as the `probability distribution'  ($pdf$) and 
specifies the probability (or probability density, for the data having continuous as
opposed to discrete values) for obtaining different values of the data, when the parameter(s)
are specified. Without this 
it is impossible to apply the likelihood (or many other) approaches. 
For example $x$ could be observations of a variable of interest within some  
range, and $f$ could be
any function such as a straight line, with gradient and intercept as parameters.
But we will start with an angular distribution
\begin{equation}
\label{pdf}
y(\cos\theta;\beta) = \frac{d\ p}{d\cos\theta} = N(1+\beta \cos^2\theta)
\end{equation}
Here $\theta$ is the angle at which a particle is observed, $dp/d\cos\theta$ is the $pdf$
specifying the probability density for observing a decay at any $\cos\theta$, $\beta$ is
the parameter we want to determine, and $N$ is the crucial nomalisation factor 
which ensures that the probability of observing a given decay at any $\cos\theta$
in the whole range from $-1$ to $+1$ is unity. In this case $N = 1/(2(1+\beta/3))$. 
The data consists of $N$ decays, with their individual observations $\cos\theta_i$.

Assuming temporarily that the value of the parameter $\beta$ is specified,
the probability density $y_1$ of observing the first decay at $\cos\theta_1$ is
\begin{equation}
y_1 = N (1+\beta \cos^2\theta_1) = 0.5 (1+\beta \cos^2\theta_1)/(1 + \beta/3),
\end{equation}
and similarly for the rest of the $N$ observations. Since the individual observations
are independent, the overall probability $P(\beta)$ of observing the complete data set
of $N$ events is given by the product of the individual probabilities
\begin{equation}
P(\beta) = \Pi y_i = \Pi \ 0.5 (1+\beta \cos^2\theta_i)/(1 + \beta/3) 
\end{equation}
We imagine that this is computed for all values of the parameter $\beta$; 
then this is known as the likelihood function ${\it L}(\beta)$.

The likelihood method then takes as the estimate of $\beta$ that value which 
maximises the likelihood. That is, it is the value which maximises (with respect to 
$\beta$) the probability density of observing the given data set. Conversely 
we rule out values of $\beta$ for which ${\it L}(\beta)$ is very small. The
uncertainty on $\beta$ is related to the width of the ${\it L}(\beta)$ 
distribution (see later). 

It is often convenient to consider the logarithm of the likelihood
\begin{equation} 
{\it l} = \ln{\it L} = \Sigma \ln y_i
\end{equation}
One reason for this is that, for a large number of observations 
some fraction could have small $y_i$. Then the likelihood, involving the product of the
$y_i$, could be very small and may underflow the computer's range for real numbers.
In contrast, {\it l} involves a sum rather than a product, and $\ln y_i$ rather than 
$y_i$, and so produces a gentler number.

\subsection{Likelihood and $pdf$}
The procedure for constructing the likelihood is first to write down the $pdf$, and then to insert into that 
expression the observed data values in order to evaluate their product, which is the likelihood. Thus both 
the $pdf$ and the likelihood involve the data $x$ and the parameter(s) $\mu$. The difference is that the $pdf$ is a function of $x$ for fixed values of $\mu$, while the likelihood is a function of $\mu$ given the fixed observed 
data $x_{obs}$.   

Thus for a Poisson distribution, the probability of observing $n$ events when the rate $\mu$ is specified is 
\begin{equation}
P(n;\mu) = e^{-\mu} \mu^n /n!
\end{equation}
and is a function of $n$, while the likelihood is
\begin{equation}
L(\mu;n) = e^{-\mu} \mu^n /n!
\end{equation}
and is a function of $\mu$ for the fixed observed number $n$.

\begin{figure}
\begin{center}
\includegraphics[width=0.75\textwidth]{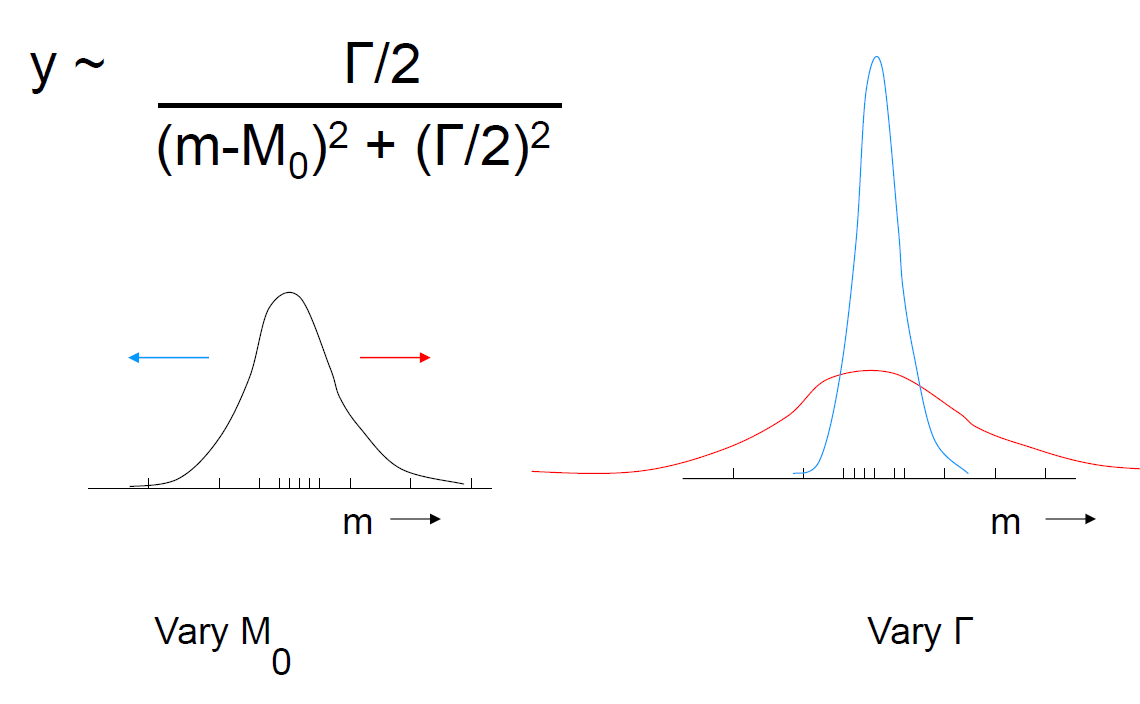}
\caption{A visual demonstration of how the maximum likelihood method gives sensible values for
the parameters, the position and width of a resonance. The bars along the $m$-axis represent the 
experimental measurements of a set of mass values $m_i$, which are to be fitted by a simple Breit-Wigner resonance
shape. In (a), the width $\Gamma$ of the resonance is kept fixed, and the mass parameter $M_0$ 
is varied. This has the effect of sliding the curve to the left or right along the $m$-axis, 
without changing its shape or height. To calculate the likelihood for a given position of the curve
we multiply all the $y(m_i)$ values; i.e. the height of the curve at each observed mass. The best value of $M_0$
is thus equivalent to finding the best location of the curve in order to maximise this product. Clearly we 
need to locate  the peak near where most of the data values are. In (b), we regard $M_0$ as constant, but vary the width. The effect of the normalisation condition then means
that the wider curve will have lower peak height and vice versa. The narrow curve suffers becase of the very small 
$y$ values for the extreme observed mass values, while wide curves do not benefit so much from the concentration of 
masses around the central value. The best value of $\Gamma$ is the result of a compromise between these two effects.}
\label{fig:L_for_Resonance}
\end{center}
\end{figure}

\subsection{Intuitive example: Location and width of peak}
We consider a 
situation where we are studying a resonant state which would result in a bump in the mass distribution of its decay particles.
We assume that the bump can be parametrised as a simple Breit-Wigner
\begin{equation}
                   y(m;M_0,\Gamma) = \frac{\Gamma/(2\pi)}{(m-M_0)^2 + (\Gamma/2)^2}
\end{equation}
where $y$ is the probability density of obtaining a mass $m$ if the location and width the state are $M_0$ and $\Gamma$,
the parameters we want to determine. It is essential that $y$ is normalised, i.e. its integral over all physical values of 
$m$ is unity; hence the normalisation factor of $\Gamma/(2\pi)$. The data consists of $n$ observations of $m$, as shown in fig. \ref{fig:L_for_Resonance}.

Assume for the moment that we know $M_0$ and $\Gamma$. Then the probability density for observing the $i^{th}$
event with mass $m_i$ is
\begin{equation}
                   y_i(M_0,\Gamma) = \frac{\Gamma/(2\pi)}{(m_i-M_0)^2 + (\Gamma/2)^2}
\end{equation}
Since the events are independent, the probability density for observing the whole data sample is
\begin{equation}
\label{product}
                   y_{all}(M_0,\Gamma) =\Pi \ \frac{\Gamma/(2\pi)}{(m_i-M_0)^2 + (\Gamma/2)^2}
\end{equation}
and this is known as the likelihood $L(M_0,\Gamma)$.  Then the best values for the parameters are taken as
the combination that maximises the probability density for the whole data sample i.e. $L(M_0,\Gamma)$. 
Parameter values for which $L$ is very small compared to its maximum value are rejected, and the uncertainties 
on the parameters are related to the width of the distribution of $L$; we will be more specific later.

 The curve in
fig. \ref{fig:L_for_Resonance}(left) shows the expected probability distribution for fixed parameter values.  The way $L$ is calculated involves
multiplying the heights of the curve at all the observed $m_i$ values. If we now consider varying $M_0$, this moves the curve bodily to the left or right without changing its shape or normalisation. So to determine the best value of $M_0$, we need to find where to locate the curve so that the product of the heights is a maximum; it is plausibe that the peak will be located where the majority of events are to be found.

Now we will consider how the optimum value of $\Gamma$ is obtained. A small $\Gamma$ results in a narrow curve, so the masses in the tail will make an even smaller contribution to the product in eqn. \ref{product}, and hence reduce the likelihood. But a large $\Gamma$ is not good, because not only is the width larger, but because of the normalisation condition, the peak height is reduced, and so the observations in the peak region make a smaller contribution to the likelihood. The optimal 
$\Gamma$ involves a trade-off between these two effects.

Of course, in finding the optimal of values of the two parameters, in general it is necessary to find the maximum of the 
likelihood as a  function of the two parameters, rather than maximising with respect to just one, and then with respect to the other and then stopping (see section \ref{More_variables}).

\subsection{Uncertainty on parameter}
With a large amount of data, the likelihood as a function of a parameter $\mu$ is 
often approximately Gaussian. In that case, ${\it l}$ is an upturned parabola. Then
the following definitions of $\sigma_\mu$, the uncertainty on $\mu_{best}$, 
yield identical answers:
\begin{itemize}
\item{The RMS of the likelihood distribution.}

\item{[$-\frac{d^2 {\it l}}{d \mu^2}]^{-1/2}$. If you remember that 
the second derivative of the log likelihood function is involved because it 
controls the width of the ${\it l}$ distribution, a mneumonic helps 
you remember the formula for $\sigma_\mu$: Since $\sigma_\mu$ must have the 
same units as $\mu$, the second derivative must appear to the power $-1/2$. But because the
log of the likelihood has a maximum, the second derivative is negative, so the minus 
sign is necessary before we take the square root.}
\item{It is the distance in $\mu$ from the maximum in order to decrease ${\it l}$  by half a unit
from its maximum value. i.e.
\begin{equation}
{\it l} (\mu_{best} + \sigma_{\mu}) = {\it l}_{max} - 0.5 
\end{equation}
}
\end{itemize}

In situations where the likelihood is not Gaussian in shape, these three definitions no longer agree.
The third one is most commonly used in that case. Now the upper and lower ends of the intervals can 
be asymmetric with respect to the central value. It is a mistake to believe that this method 
provides intervals which have a $68\%$ chance of containing the true value of the 
parameter\footnote{Unfortunately, this incorrect statement occurs in my book\cite{LL_book}. It is 
corrected in a separate update\cite{LL_book_update}.}.

Symmetric uncertainties are easier to work with than asymmetric ones. It is thus sometimes better to quote the 
uncertainty on a function of the first variable you think of. For example, for a charged particle in a magnetic field,
the reciprocal of the momentum has a nearly symmetric uncertainty. Especially for high
momentum tracks, the upper uncertainty on the momentum can be much larger than the lower one 
e.g. $1.0\ ^{+1.5}_{-0.4}$ TeV.

\begin{figure}
\begin{center}
\includegraphics[width=1.0\textwidth]{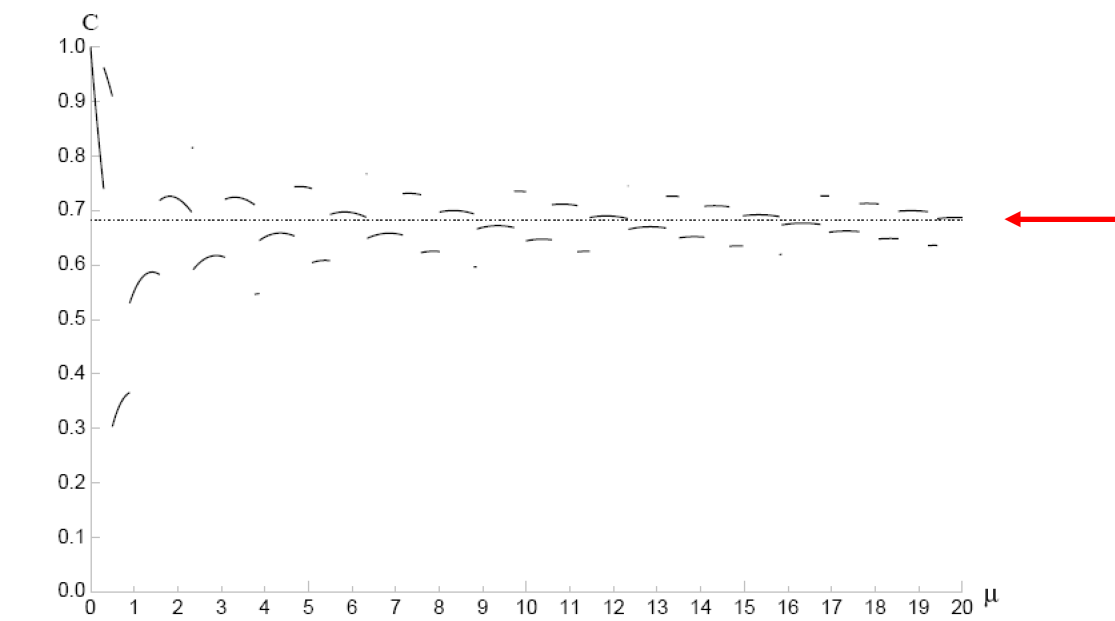}
\caption{Coverage C for Poisson parameter intervals, as determined by the $\Delta(log(L)) =0.5$ rule. 
Repeated trials (all using the same Poisson parameter $\mu$) yield different values of $n$, each 
resulting in its own range $\mu_l$ to $\mu_u$ for $\mu$; then  C is the fraction of trials that give ranges which 
include the chosen value of $\mu$ for the trials. The coverage C varies with $\mu$, and has 
discontinuities because the data $n$ can take on only discrete integer values. For large $\mu$, C 
seems to approach the expected 0.68, shown by the arrow, but for small $\mu$, the coverage takes on 
values between $30\%$ and $100\%$.   
\label{fig:PoissonCoverage}}
\end{center}
\end{figure}

\subsection{Coverage}
An important feature of any statistical method for estimating a range for some parameter $\mu$ at a 
specified confidence level $\alpha$ is its coverage $C$. If the procedure is applied many times, 
these ranges will vary because of statistical fluctuations in the observed data. Then $C$ is defined as
the fraction of ranges which contain the true value $\mu_{true}$; it can vary with $\mu_{true}$. 

It is very
important to realise that coverage is a property of the {\bf statistical procedure} and does not apply
to your particular measurement. An ideal plot of coverage as a function of $\mu$ would have $C$ constant 
at its nominal value $\alpha$. For a Poisson counting experiment, figure \ref{fig:PoissonCoverage} shows $C$ as a 
function of the Poisson parameter $\mu$, when the observed  number of counts $n$ is used to determine a range 
for $\mu$ via the change in log-likelihood being 0.5. The coverage is far from constant at small $\mu$.
 If $C$ is smaller than $\alpha$, this is known as undercoverage. Certainly frequentists would regard this 
as unfortunate; it means that people reading an article containing parameters determined this way are 
likely to place  more than justified reliance on the quoted range. Methods using the Neyman construction 
to determine parameter ranges by construction do not have undercoverage.     

Coverage involves a statement about $Prob[\mu_l \leq \mu_{true} \leq \mu_u]$. This is to be interpreted as a
probability statement about how often the ranges $\mu_l$ to $\mu_u$ contain the (unknown but constant) true
value $\mu_{true}$. This is a frequentist statement; Bayesians do not want to consider the ensemble of possible
results if the measurement procedure were to be repeated. Thus Bayesians would regard the statement
about $Prob[\mu_l \leq \mu_{true} \leq \mu_u]$  as describing what fraction of their estimated 
posterior probability density for $\mu_{true}$ would be 
between the fixed values $\mu_l$ and $\mu_u$, derived from their actual measurement.

\subsection{More than one parameter}
\label{More_variables}

For the case of just one parameter $\mu$, the likelihood best estimate $\hat{\mu}$ is given 
by the value of $\mu$ which maximises $L$. Its uncertainty $\sigma_\mu$ is determined either from 
\begin{equation}
\label{error}
1/\sigma_\mu^2 = -d^2\ln L/d\mu^2 ;
\end{equation} 
of by finding how far $\hat{\mu}$ would have to be changed in order to reduce $\ln L$ by 0.5.

When we have two or more parameters $\beta_i$ the rule for finding the best estimates $\hat{\beta_i}$
is still to maximise $L$.
For the uncertainties and their correlations, the generalisation of equation \ref{error} is to construct
the inverse covariance matrix ${\bf M}$, whose elements are given by
\begin{equation}
M_{ij} = -\frac{\partial^2 \ln L} {\partial \beta_i\ \partial \beta_j} 
\end{equation} 
Then the inverse of $\bf{M}$ is the covariance matrix, whose diagonal elements are the variances of $\beta_i$,
and whose off-diagonal ones are the covariances.

Alternatively (and more common in practice), the uncertainty on a specific $\beta_j$ can be obtained 
by using the profile likelihood $L_{prof}(\beta_j)$.
This is the likelihood as a function of the specific $\beta_j$, where for each value of $\beta_j,\ L$ has been remaximised
with respect to all the other $\beta$. Then $L_{prof}(\beta_j)$ is used with the `reduce $\ln L_{prof}$ = 0.5' rule
to obtain the uncertainty on $\beta_j$. This is equivalent to determining the contour in $\beta$-space where 
$\ln L = \ln L_{max} - 0.5$, and finding the values $\beta_{j,1}$ and $\beta_{j,2}$ on the contour which are   
furthest from $\hat{\beta_j.}$ Then the (probably asymmetric) upper and lower uncertainties on $\beta_j$ are 
given by $\beta_{j,2}- \hat{\beta_j}$ and $\hat{\beta_j} - \beta_{j,1}$ respectively.

Because these are likelihood methods of obtaining the intervals, these estimates of uncertainities provide only
{\bf nominal} regions of 68\% coverage for each parameter; the {\bf actual} coverage can differ from this. Furthermore, 
the region within 
the contour described in the previous paragraph for the multidimensional $\beta$ space will have less than 68\% nominal
overage. To achieve that, the $`0.5'$ in the rule for how much $\ln L$ has to be reduced from its maximum    
 must be replaced by a larger number, whose value depends on the dimensionality of $\beta$.

 \section{Worked example: Lifetime determination}
\label{sec:lifetime_example}
Here we consider an experiment which has resulted in $N$ observed decay times $t_i$ of a particle
whose lifetime $\tau$ we want to determine. The probability density for observing a decay at time $t$ 
is 
\begin{equation}
\label{exp}
p(t;\tau) = (1/\tau) \  e^{-t/\tau}
\end{equation}  
Note the essential normalisation factor  $1/\tau$; without this the likelihood method does not work.

It should be realised that realistic situations are more complicated than this. For example, we ignore
the possibility of backgrounds, time resolution which smears the expected values of $t$, acceptance or 
efficiency effects which vary with $t$, etc., but this enables us to estimate $\tau$ and its uncertainty
$\sigma_{\tau}$ analytically. In real practical cases, it is almost always necessary to calculate the 
likelihood as a function of $\tau$ numerically. 
 
From equation \ref{exp} we calculate the log-likelihood as
\begin{equation}
\ln L(\tau) = \ln[\Pi\ (1/\tau) e^{-t_i/\tau}] \ \ = \ \ \Sigma (-\ln \tau - t_i/\tau)
\end{equation}
Differentiating $\ln L(\tau)$ with respect to $\tau$ and setting the derivative to zero then yields
\begin{equation}
\tau = \Sigma t_i/N
\end{equation}
This equation has an appealing feature, as it can be read as ``The mean lifetime is equal to the mean lifetime",
which sounds as if it must be true. However, what it really says is not quite so trivial: ``Our
best estimate of the lifetime parameter $\tau$ is equal to the mean of the $N$ observed decay times in our 
experiment."

We next calculate $\sigma_{\tau}$ from the second derivative of $\ln L$, and obtain 
\begin{equation}
\sigma_{\tau} = \tau/\sqrt N
\end{equation}
This exhibits a common feature that the uncertainty of our parameter estimate decreases as $1/\sqrt N$ as we 
collect more and more data. However, a potential problem arises from the fact that our estimated uncertainty
is proportional to our estimate of the parameter. This is relevant if we are trying to combine different experimental
results on the lifetime of a particle. For combining procedures which weight each result by $1/\sigma^2$, a 
measurement where the fluctuations in the observed times  result in a low estimate of $\tau$
will tend to be over-weighted (compare the section on `Combining Experiments' in Lecture 1), 
and so the weighted average would be biassed 
downwards. This shows that it is better to combine different experiments at the data level, rather than 
simply trying to use their results.

One final point to note about our simplified example is that the likelihood $L(\tau)$ depends on the 
observations only via the {\bf sum} of the times $\Sigma t_i$ i.e. their {\bf distribution} is 
irrelevant. Thus the likelihood distributions for two experiments having the same number of events and the 
same sum of observed decay times, but with one having the decay times consistent with an exponential 
distribution and the other having something completely different (e.g. all decays occur at the same time), 
would have identical likelihood
functions. This provides an example of the fact that the unbinned likelihood function does not in general provide
useful information on Goodness of Fit.


\section{Conclusions}
Jut as it is impossible to learn to play the violin without ever picking it up and spending hours 
actually using it, it is important to realise that one does not learn how to apply Statistics 
merely by listening to lectures. It is really important to work through examples and actual analyses, and 
to discover more about the topics.

There are many resources that are available to help you. First there are textbooks written by Particle 
Physicists\cite{books}, which address the statistical problems that occur in Particle Physics, and which use a language
which is easier for other Particle Physicists to understand.

The large experimental collaborations have Statistics Committees, whose web-sites contain lots of 
useful statistical information. That of CDF\cite{CDF_St_Cttee} 
is most accessible to Physicists from other experiments.             

The Particle Data Book\cite{PDG} contains short sections on Probability, Statistics and Monte Carlo
simulation. These are concise, and are useful reminders of things you already know. It is harder to
use them instead of lengthier articles and textbooks in order to understand a new topic.

If in the course of an analysis you come upon some interesting statistical problem that you do not immediately 
know how to solve,  you might be tempted to invent your own method of how to overcome the problem. This
can amount to reinventing the wheel. It is a good idea to try to see if statisticians (or even Particle Physicists) have 
already dealt with this topic, as it is far preferable to use their circular wheels, rather than your own hexagonal one.

Finally I wish you the best of luck with the statistical analyses of your data.


\clearpage



\begin{thebibliography}{99}
\bibitem{Lecture3} 'Bayes and Frequentism: The return of an old controversy', Louis Lyons, under Recommendations section on CDF statistics committee page \url{https://www-cdf.fnal.gov/physics/statistics/} (2002) 
\bibitem{Lecture4}  'Statistical issues in searches for New Physics', Louis Lyons, under Notes section on CDF statistics committee page \url{https://www-cdf.fnal.gov/physics/statistics/} (2002) 
\bibitem{Lecture5} `Learning to Love the Error Matrix'. A video and the slides of an earlier version of the lecture can be found at
\url{http://vmsstreamer1.fnal.gov/VMS_Site_03/Lectures/AcademicLectures/presentations/040803Lyons.pdf    }.


\bibitem{LL_book} Louis Lyons, `Statistics for nuclear and particle physicists',  Cambridge University Press (1986).
\bibitem{LL_book_update} Louis Lyons, `Statistics for Nuclear and Particle Physicists: an Update',
\url{http://www-cdf.fnal.gov/physics/statistics/notes/Errata2.pdf } (2009). 
\bibitem{DAgostino} Ralph B. D'Agostino and Michael A. Stephens, `Goodness of Fit Techniques', CRC Press (1986).
\bibitem{Coverage}Joel G. Heinrich, `Coverage of Error Bars for Poisson Data', \url{http://www-cdf.fnal.gov/physics/statistics/notes/cdf6438_coverage.pdf}.

\bibitem{books}R.J. Barlow `Statistics' (Wiley,1989);\\
O. Behnke et al. (editors),`Data Analysis in High Energy Physics: a Practical Guide to Statistical
Methods' (Wiley. 2013);\\
G. Cowan, `Statistical Data Analysis' (OUP, 1998);\\
F. James, `Statistical Methods in Experimental Physics', (World Scientific, 2006);\\
L. Lista, `Statistical Methods for Data Analysis in Particle Physics', (Springer, 2015);\\
L. Lyons, `Statistics for Nuclear and Particle Physicists' (CUP, 1986);\\
B. Roe, `Probability and Statistics in Experimental Physics' (Springer, 1992)
\bibitem{CDF_St_Cttee} CDF Statistics Committee web-site, \url{https://www-cdf.fnal.gov/physics/statistics/}
\bibitem{PDG} G. Cowan in Particle Data Group - 2016 Review',
\url{ http://pdg.lbl.gov/2016/reviews/rpp2016-rev-statistics.pdf};\\
\url{http://pdg.lbl.gov/2016/reviews/rpp2016-rev-probability.pdf}; and \\ 
\url{http://pdg.lbl.gov/2016/reviews/rpp2016-rev-monte-carlo-techniques.pdf}
\end{thebibliography}
\end{document}